# Grain Size and Texture of $Cu_2ZnSnS_4$ Thin Films Synthesized by Cosputtering Binary Sulfides and Annealing: Effects of Processing Conditions and Sodium

W.M. HLAING OO,[1] J.L. JOHNSON,[2] A. BHATIA,[1] E.A. LUND,[3] M.M. NOWELL,[4] and M.A. SCARPULLA[1,2,5]

[1] *Department of Materials Science and Engineering, University of Utah, Salt Lake City, UT, USA.*

[2] *Department of Electrical and Computer Engineering, University of Utah, Salt Lake City, UT, USA.*

[3] *Department of Chemical Engineering, University of Utah, Salt Lake City, UT, USA.*

[4] *EDAX-TSL, 392 East 12300 South Suite H., Draper, UT, USA.*

[5] *e-mail: scarpulla@eng.utah.edu*

**ABSTRACT**

We investigate the synthesis of kesterite $Cu_2ZnSnS_4$ (CZTS) polycrystalline thin films using cosputtering from binary sulfide targets followed by annealing in sulfur vapor at 500 °C to 650 °C. The films are the kesterite CZTS phase as indicated by x-ray diffraction, Raman scattering, and optical absorption measurements. The films exhibit (112) fiber texture and preferred low-angle and $\Sigma 3$ grain boundary populations which have been demonstrated to reduce recombination in $Cu(In,Ga)Se_2$ and CdTe films. The grain growth kinetics are investigated as functions of temperature and the addition of Na. Significantly, lateral grain sizes above 1 μm are demonstrated for samples grown on Na-free glass, demonstrating the feasibility for CZTS growth on substrates other than soda lime glass.

_________________________________________________________________________________________________

**INTRODUCTION**

The kesterite phase of quaternary $Cu_2ZnSnS_4$ (CZTS) has attracted interest as a possible candidate for earth-abundant photovoltaic (PV) absorber layers which could be used to reduce materials costs for a large-scale PV industry producing >150 $GW_P$/ year of thin-film panels.[1–9] CZTS has very similar structural and optoelectronic properties to the chalcopyrite $Cu(In,Ga)(S,Se)_2$ (CIGSSe) alloys, which are the thin-film absorber materials used in current thin-film cells with record laboratory efficiency.[10] Research and development on CZTS is progressing at a rate comparable to that seen in the early years of CIGSSe research, and the current record device efficiency for a soda lime glass/Mo/ CZTS/CdS/i-ZnO/ZnO:Al cell with CZTS synthesized by thermal evaporation is 6.8%.[9,11] To date, cells based on the Se analog $Cu_2ZnSnSe_4$ (CZTSe) have demonstrated higher efficiencies up to 9.7%.[12] As for all existing thin-film PV materials, device performance is limited by defects and imperfections within the materials and at interfaces between materials; thus it is important to fully understand and control these defects in order to evaluate the practicality of using CZTS in thin-film solar cells manufactured at large scales.

In this work we report on the synthesis and properties of CZTS thin films synthesized by ambient-temperature cosputtering from binary sulfide targets followed by long-term (2 h) annealing in sulfur vapor in the 500°C to 650°C range to promote compositional homogenization, phase formation, grain growth, and low point and interface defect densities. CZTS thin films have been fabricated by many synthesis routes similar to this one, including sulfurization of electrodeposited,[13] electron-beamevaporated,[14] and radiofrequency (RF)-sputtered[15] precursors. We have previously reported our preliminary work on CZTS formed by sulfurizing layers cosputtered from binary sulfides,[16,17] and other groups have also attempted related sputtering routes.[9,18,19] We demonstrate that this synthesis route results in thin polycrystalline films with good compositional uniformity, grain size up to a few microns, and highly (112) textured films with $\Sigma 3$ grain boundaries preferentially in plane. However, Raman scattering suggests the possibility of the coexistence of ZnS as a second phase. We are currently exploring the defect properties of these films as well as refining the synthesis conditions to yield higher-quality thin films.

**EXPERIMENTAL PROCEDURES**

Mo was sputter-deposited in-house using deposition processes we have demonstrated to result in very low residual stress of <23 MPa and strong (110) texture.[17] Thin-film CZTS samples were deposited on Mo-coated soda lime glass (SLG) and boroaluminosilicate glass (BSG) substrates by simultaneous RF sputtering from $Cu_2S$, $ZnS$, and $SnS_2$ targets of >99.5% purity (Plasmaterials, Livermore, CA). CZTS deposition was carried out in a load-locked sputtering chamber pumped by a turbomolecular pump backed by an oil-free scroll roughing pump. Additionally, liquid-nitrogen-cooled tubes inside the chamber cryopump gases such as water, resulting in clean-chamber base pressures of $2 \times 10^{-7}$ Torr or lower. Before CZTS deposition the Mo-coated substrates were plasma cleaned in the deposition chamber for 10 min at 100 W RF bias with 20 sccm argon at 5 mTorr; this results in lower back-contact resistance, presumably because it removes the oxidized layer from the Mo.[17] Argon is injected at 20 sccm close to each 75-mm-diameter sputtering target in the Lesker Torus° sputtering sources. The sputtering chamber pressure is regulated in the 1 mTorr to 10 mTorr range using a variable butterfly valve before the turbomolecular pump. The RF deposition powers resulting in near-stoichiometric as-deposited CZTS thin films were close to 110 W for $Cu_2S$, 90 W for ZnS, and 45 W for $SnS_2$. The target–substrate distance was 16.5 cm, and the sample was rotated during deposition. These parameters result in film growth rate of 0.844 μm/h and ±0.5% (relative) fluctuations in composition and thickness across the 75 mm x 75 mm deposition area. The substrate temperature is not actively controlled and may reach up to 125°C during long depositions.

Following deposition, the sputtered films were annealed in a tube furnace under sulfur vapor to form the CZTS phase, grow grains, and further homogenize the composition. During the annealing the sulfur vapor was supplied from heated sulfur powder in a boat inside the tube but outside of the hot zone of the furnace. The temperature of the boat was kept at 170°C, and was always maintained for the ramp-up, annealing, and cooling times while the samples were above ~150°C. The sample temperature was ramped up over 2 h to the annealing temperatures of 550°C, 600°C, and 650°C where it was held for 2 h and then allowed to cool slowly overnight (~20 h) before removal from the furnace. There was always excess S left in the boat at the end of the annealing. Samples for optical transmission and electrical measurements were deposited directly on SLG without Mo.

The properties of the films were characterized by Hall-effect measurements, optical absorption, x-ray diffraction (XRD), Raman spectroscopy, energy-dispersive x-ray spectroscopy (EDX), inductively coupled plasma mass spectrometry (ICPMS), electron backscatter diffraction (EBSD), and scanning electron microscopy (SEM). Hall measurements were performed at 1 T on square samples in van der Pauw geometry using sputtered Ohmic Mo contacts on each corner of the sample. The optical absorption spectra were taken using a 0.5 m monochromator using a Xe-Hg light source and a Si detector. Raman spectra were taken using 532 nm laser excitation. The θ–2θ XRD patterns were collected on a diffractometer having a Goebel mirror, a divergence slit of 1°, an antiscatter slit of ½°, and a receiver slit of ¼°.

The polycrystalline texture of the samples was examined using EBSD after using a focused $Ga^+$ ion beam (FIB) to mill and polish a surface through the sample thickness at grazing incidence. To do this, samples were first mounted on a 38° pretilted holder, aligned with the beam, and then tilted 1.5° from this position. The sample was milled at 5 kV beam energy and 1.5 nA beam current, resulting in a wedged surface through the thickness of the film. The samples were then transferred to a field emission SEM equipped with an EBSD detector. The samples were oriented such that the milled analysis plane was tilted 15° away from the incident electron beam and towards the EBSD detector. This was done by aligning the analysis plane with the incident electron beam and then changing the stage tilt value by 15°. This places the analysis plane in a typical geometry for EBSD collection with a 75° tilt relative to a standard nontilted position. The EBSD patterns were collected at 20 kV acceleration voltage, beam current of 3 nA, a 14 mm working distance, and lateral step size of 35 nm. Grains were determined from contiguous EBSD points having misorientation angle<5°; i.e., a cutoff of <5° misorientation was used to define contiguous grains. The grain boundary misorientation distribution was determined by binning the misorientation data from 5° to 65° into 3° bins and then plotting the resulting histogram. Because the c/2a ratio is closer to 1 than can be distinguished by EBSD,[20] the patterns were indexed using the zincblende unit cell. This procedure will map grain boundary misorientation angles between the maxima for zincblende (62.8°) and for tetragonal (90°) onto the interval 0° to 27.2°.

**RESULTS AND DISCUSSION**

The compositions of the samples were examined in the scanning electron microscope using EDX with an expected uncertainty of a few at.% and with ICPMS with higher accuracy and precision as shown in Table 1. This paper reports results from samples cosputtered in batches numbered 35 and 37 from our laboratory. The as-deposited compositions were designed such that the metal ratios were close to stoichiometric (as seen especially in the ICPMS data) with the primary deviation being a deficiency of sulfur which is later diffused into the films in the annealing step. For both sample batches, the as-deposited Sn content was very close to stoichiometric. The #35 samples were slightly Cu-rich and correspondingly Zn-poor, while the #37 samples were slightly Cu-poor and correspondingly Zn-rich. After annealing at 550°C or 600°C, there is a general trend towards some loss of Sn (barely above the experimental precision of EDX measurements). Also, since EDX will sample the entire film depth for these samples which are 0.5 μm to 1 μm thick, the compositions reported in the table are calculated including the effects of x-ray absorption through the film thickness assuming a uniform depth distribution of each element. ICPMS is performed after dissolving the entire film. Thus, the small but significant Cu excess and Sn loss indicated by EDX for the SLG sample annealed at 600°C, taken with the observations from ICPMS that the average compositions did not change so drastically, does seem to suggest some segregation of the Cu towards the surface for this Na-containing sample.

| Glass | Annealing | Sample ID | EDX | | | | ICPMS | | | |
|---|---|---|---|---|---|---|---|---|---|---|
| | | | Cu/Metals (%) | Zn/Metals (%) | Sn/Metals (%) | S/Total (%) | Cu/Metals (%) | Zn/Metals (%) | Sn/Metals (%) | S/Total (%) |
| SLG | As-Deposited | #35 | 51 | 24 | 25 | 47 | 50.4 | 24.2 | 25.5 | 47.0 |
| | As-Deposited | #37 | 49 | 25 | 26 | 48 | - | - | - | - |
| | 550 °C | #35 | 48 | 28 | 24 | 49 | - | - | - | - |
| | 550 °C | #37 | 47 | 26 | 26 | 51 | - | - | - | - |
| | 600 °C | #35 | 54 | 25 | 21 | 50 | 49.2 | 27.7 | 23.1 | 49.0 |
| BSG | As-Deposited | #35 | 51 | 24 | 25 | 47 | 50.0 | 24.7 | 25.3 | 45.0 |
| | As-Deposited | #37 | 48 | 27 | 25 | 48 | - | - | - | - |
| | 550 °C | #37 | 48 | 28 | 25 | 48 | - | - | - | - |
| | 600 °C | #35 | 51 | 25 | 23 | 47 | 50.7 | 24.6 | 24.6 | 43.0 |
| | 600 °C | #37 | 49 | 26 | 25 | 48 | - | - | - | - |

**Table 1** - Compositions of as-deposited and annealed samples identically processed on SLG and BSG as measured using EDX and ICPMS. Note that the Cu, Zn, and Sn compositions are normalized to the total cation composition [Cu]+[Zn]+[Sn] while [S] is normalized to the total composition [Cu]+[Zn]+[Sn]+[S].

The indicated sulfur composition for annealed samples on Mo is slightly lower than stoichiometric (47% to 48%), but it should be noted that otherwise identical samples processed on bare glass yield EDX-measured S composition approximately 2% to 2.5% higher. This discrepancy results from EDX peak overlap between the S K and Mo L peaks, and we therefore place more confidence in the higher, very nearly stoichiometric measurements made on samples without Mo. Note that the measurements of S using ICPMS are systematically lower for all samples; this is attributed to the fact that the samples were dissolved in a solution of nitric acid, from which volatilization of $H_2S$ and possibly other S-containing gasses is expected. Thus, the measurements of S using ICPMS are discounted for this work.

Hall-effect and resistivity measurements were made on companion samples deposited and annealed on SLG and BSG with no Mo. All samples showed p-type conductivity, however truly accurate values of carrier concentration were not obtainable since the indicated Hall mobility is extremely small (on the order of 0.1 $cm^2$/V·s to 1 $cm^2$/ V·s), similar to reports from other groups.[21] This is consistent with the carrier concentrations estimated from the resistivity assuming mobilities of 1 $cm^2$/ V·s to 10 $cm^2$/ V·s (characteristic of CIGSSe films) ranging from $10^{18}$ $cm^{-3}$ to $10^{19}$ $cm^{-3}$. These large carrier concentrations are caused by the measured departures from ideal stoichiometry.

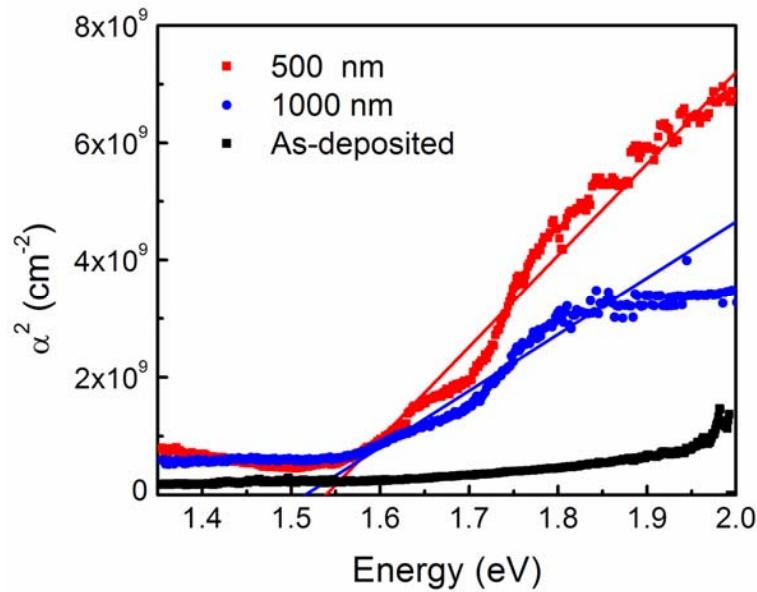

**Figure 1** - Characteristic optical absorption data for two CZTS thin films deposited directly on SLG and annealed at 600°C but having different thicknesses.

Figure 1 shows the optical absorption spectra of an as-deposited film and two CZTS films of different thicknesses annealed at 600°C. The as-deposited sample does not show a clear direct bandgap absorption feature, while the data for the two annealed samples indicate bandgap energies between 1.5 eV and 1.55 eV and a magnitude of the absorption coefficient on the order of $10^4$ $cm^{-1}$. These are consistent with other reports for CZTS.[21,22] We may also estimate the magnitudes of the absorption coefficient from the slope of the absorption data in the high-energy regime for the two samples having different thickness. The fact that the absorption continues to increase for energies above 1.8 eV for the

500-nm-thick sample indicates that it is still optically thin in this regime, while the fact that the 1-μm-thick sample's absorption is constant in this regime indicates that its thickness exceeds approximately two absorption lengths. Thus, we may conclude that the absorption coefficient exceeds approximately 5 x $10^4$ cm$^{-1}$ for photon energies above 1.8 eV for these CZTS samples.

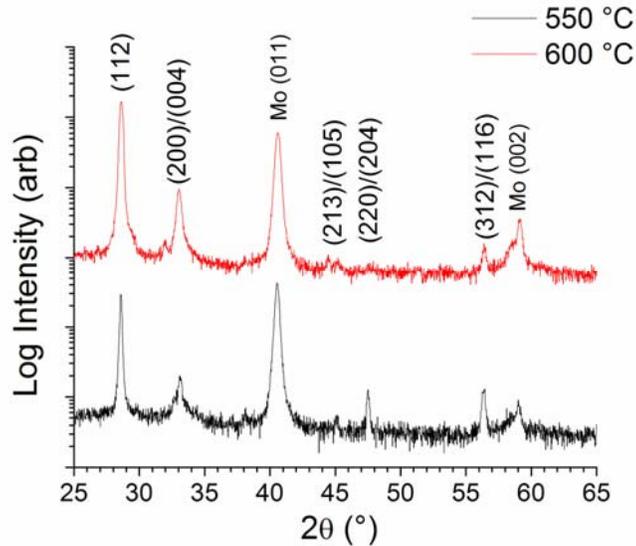

**Figure 2** - X-ray diffraction data from CZTS samples on Mo-coated SLG annealed at (a) 550°C and (b) 600°C.

Figure 2 displays symmetric θ–2θ XRD data collected from two representative samples annealed at 550°C and 600°C. All of the diffraction peaks besides the Mo (011) at 40.5° can be indexed to tetragonal $Cu_2ZnSnS_4$. The increased intensity of the (112) peak relative to other CZTS peaks seen for the sample annealed at 600°C indicates stronger (112) fiber texture for higher-temperature annealing. Because of their nearly identical lattice constants, the XRD patterns for kesterite and stannite $Cu_2ZnSnS_4$, $Cu_2SnS_3$, and ZnS are quite similar and cannot be distinguished very reliably. Therefore, we measured Raman spectroscopy to attempt detection of such subphases.

The phonon spectra of the CZTS-related phases mentioned above are distinct because of the different bonding configurations in each one. Figure 3 presents Raman spectra from the samples from Fig. 2. The strong peak near 337 cm$^{-1}$ is attributed to the kesterite CZTS phase.[11,19,23,24] However, it is clearly asymmetric towards higher wavenumber (which is incompatible with phonon confinement effects), indicating that two peaks may be overlapping. Two other smaller peaks are seen at 256 cm$^{-1}$ and 286 cm$^{-1}$. Therefore, the data were fitted with four Lorentzian components. As seen in Fig. 3, the two fitted peaks at 286 cm$^{-1}$ and 336 cm$^{-1}$ have similar peak width and are attributed to the kesterite $Cu_2ZnSnS_4$ phase. As pointed out by Fernandes et al.,[23] the small peak at 256 cm$^{-1}$ which has a larger peak width in these samples may arise from convolutions of peaks corresponding to $Sn_2S_3$, CZTS, $Cu_{2-x}S$, and ZnS and is therefore not helpful in phase identification. The other strong peak at 349 cm$^{-1}$ to 350 cm$^{-1}$ is believed to be from cubic ZnS. The peak widths of this and the 256 cm$^{-1}$ peak are very similar, suggesting the possibility of a common origin. The chemical composition data demonstrates that both samples are slightly Zn-rich (approximately 14 at.%, compared with 12.5 at.% for stoichiometric $Cu_2ZnSnS_4$), making the formation of some secondary ZnS possible.

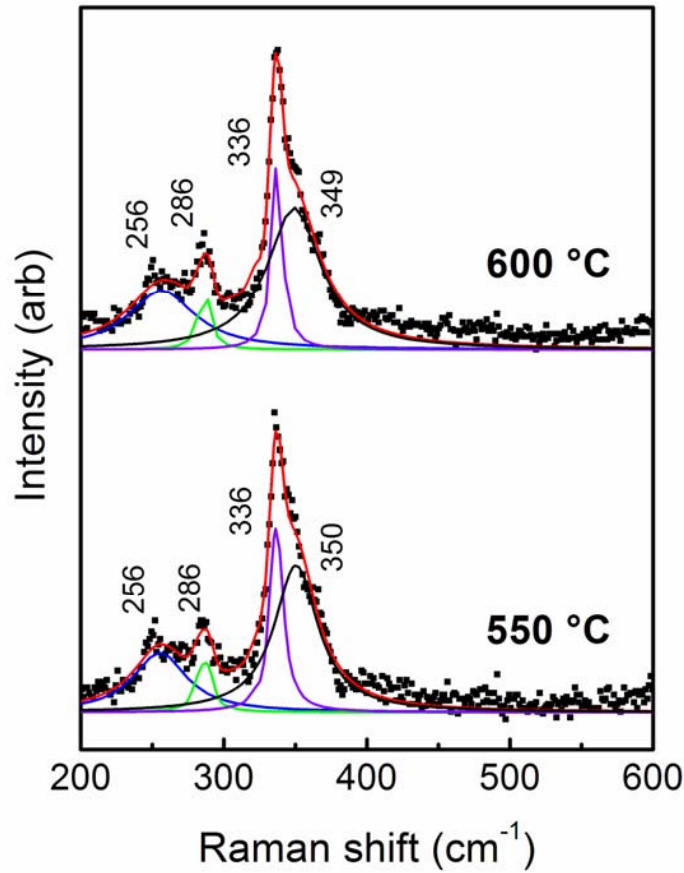

**Figure 3** - Raman spectra from samples on Mo-coated SLG annealed at (a) 550°C and (b) 600°C. Solid lines show the results of peak fitting using Lorentzian line shapes as discussed in the text, with red being the sum of the four fitted peaks.

The average lateral grain sizes of the films were determined for different annealing conditions from SEM imaging. As-deposited films exhibit lateral grain sizes in the 40 nm to 60 nm range and are equiaxed. After annealing at lower temperatures (550°C) grains tend to be columnar in shape and extend from the top to the bottom of the film. Grain shape for samples annealed at higher temperatures (600°C to 650°C) resulted in more spherical geometries. Figure 4 shows the average lateral grain sizes determined from different samples annealed at different temperatures. It can be seen that the lateral grain sizes increase with annealing temperature for samples on both BSG and SLG, but the increase in grain size for samples on SLG is more pronounced at all annealing temperatures compared with those on BSG. This appears to be similar to the case for CIGSe, where diffusion of Na from the SLG is associated with larger grain sizes (as well as with beneficial electronic effects) at elevated processing temperatures.[25–29] The hypothesis of Na diffusion during annealing is supported by the fact that we do not observe significant amounts of Na in as-deposited films on SLG using EDX.

To verify the role of sodium, Na was intentionally introduced to an as-sputtered CZTS sample on BSG (which contains no Na). Half of the sample was dipped in an aqueous solution of Na₂S which was then dried in air before the sample underwent the same annealing procedure at 600°C as the SLG samples described above. Figure 5 shows SEM images of two spots on the sample in the regions with and without Na₂S exposure. The grain size on the side exposed to Na₂S is in the 1 µm to 4 µm range while grain growth on the unexposed side is hardly visible at the same magnification. It should be noted that significant porosity developed on the side of the sample exposed to Na₂S (at an uncontrolled total amount of Na₂S deposited from solution). This may indicate that larger grains and/or very Na-rich conditions may be more susceptible to evaporation of CZTS or some of its components, especially from grain boundaries. These results clearly demonstrate that Na plays a similar role in CZTS grain growth as it does in CIGSe.

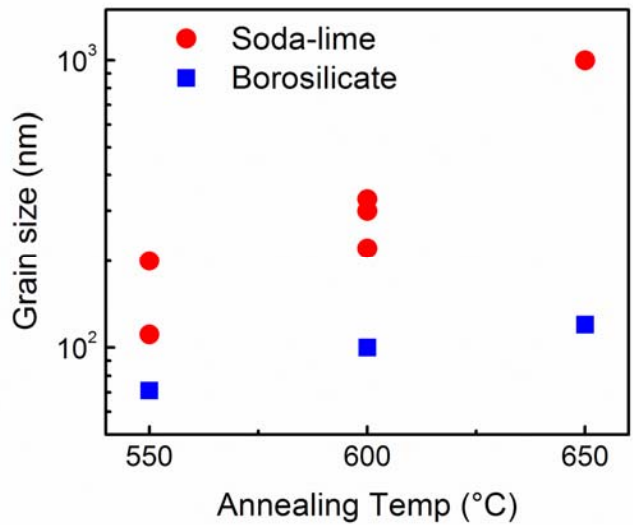

**Figure 4** - Grain size versus annealing temperature for CZTS films annealed on identically Mo-coated Na-containing (SLG) and Na-free (BSG) glass.

Using EBSD we analyzed the polycrystalline texture and orientational distributions grain boundaries of a 1250-nm-thick CZTS sample on SLG annealed at 600°C. Figure 6a shows an orientation map of the sample as collected along the FIB-milled cut through the thickness of the film. Colors in this large-area image are assigned randomly to grains having different crystallographic axes aligned out of plane. Black areas correspond to points that gave diffraction patterns that were not definitively indexable; for example, two near-surface height nonuniformities (probably pits) are seen as elongated black regions in the upper portion of the image. As can be seen, a majority of the grains are dark blue in color, corresponding to having their ⟨112⟩ directions oriented out of plane. To quantify the polycrystalline texture, pole figures were plotted from the collected data. Figure 6b shows a pole figure for the ⟨112⟩ directions, indicating that the film has significant fiber texture with the ⟨112⟩ out-of-plane direction occurring approximately at six times the rate for random crystallite orientation. A ring of slightly enhanced orientation probability occurs centered at 70.5° from the normal (the angle between ⟨112⟩ directions for a tetragonal unit cell with c/2a = 1, a good approximation for kesterite CZTS). Two pockets of orientation probability near two times random appear at azimuthal separation near 120°, indicating that there is also some weak in-plane orientation correlation between grains (in this plot a ⟨112⟩-oriented single crystal of CZTS would show a point at the center of the plot plus three points at 70.5° from normal, spaced 120° from each other azimuthally). Also from this image, an average grain size of 350 nm is obtained for the sample and it is noted that the lateral grain size does not change with depth through the film. Note that this grain size is significantly smaller than the size obtained for the sample treated with $Na_2S$, indicating that the concentration of Na (in addition to temperature) likely plays a role in grain growth kinetics. It is also apparent that, under the annealing conditions used, Na was able to diffuse sufficiently to have uniform effects through the depth of the film.

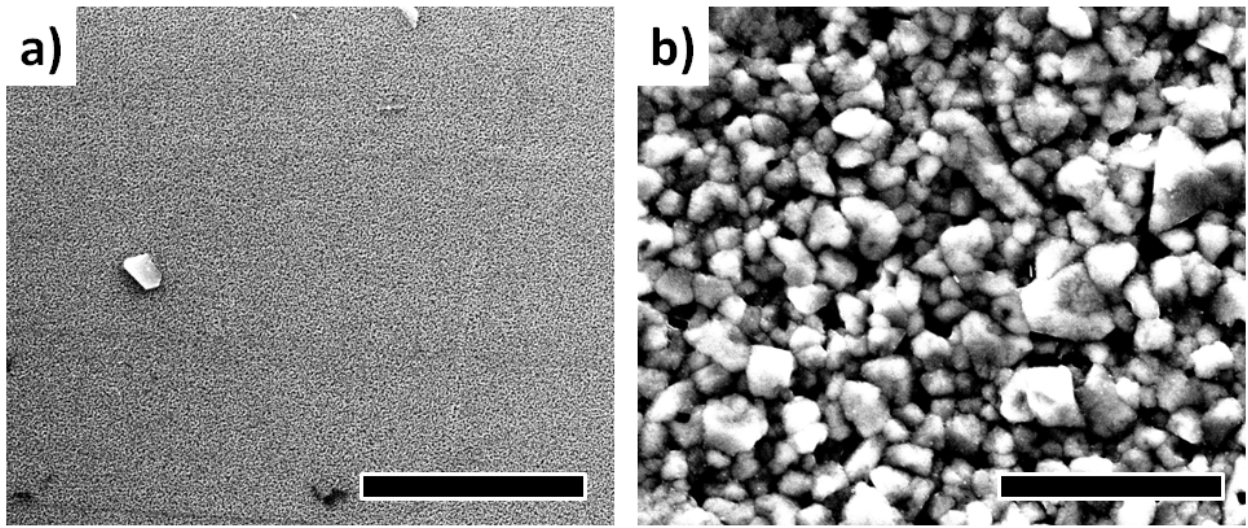

**Figure 5 -** SEM images from two regions of a CZTS film deposited on an Na-free BSG substrate and annealed in a sulfur atmosphere at 600°C for 2 h (a) as-deposited and (b) coated with $Na_2S$ from aqueous solution. The scale bars are 10 μm for both figures. Grains larger than 1 μm are formed for the Na-exposed sample, however some smaller grains and porosity are also present.

Experimental observation in CIGSe and CdTe films suggests that nonrandom polycrystalline film texture can lower the overall recombination rate in the material by the formation of special grain boundaries.[30–33] These special grain boundaries are termed coincident site lattice (CSL) boundaries and are characterized by their Σ value, which relates the unit cell volume of the CSL to that of the material crystal lattice. Low Σ values indicate strong congruence between the lattices of the two grains along the grain boundary, therefore generally corresponding to lower electronic defect density, resulting in lower recombination and better electrical transport. Figure 6c shows the distribution of grain boundaries in the image shown in Fig. 6a as a function of their misorientation angle. The (Mackenzie) curve for random grains of a cubic material is shown for comparison. As can be seen, the 60° misoriented twin boundaries (Σ3 twins) are the largest fraction of all grain boundaries and significantly exceed the random prediction. There is also some enrichment of grain boundaries misoriented by less than approximately 20°, however we consider these data to be only marginally higher than the random curve. Because of the aliasing of some of the high angle misorientations allowed for tetragonal crystals into this range by the cubic indexing procedure, we do not comment further on boundaries having misorientation angles in this range. It is clear and unambiguous that the population of Σ3 twin boundaries is significantly enhanced in the film, indicating that in-plane rotational correlations are superimposed on the overall out-of-plane (112) fiber texture. It remains to be seen if these orientation correlations indeed result in better charge transport across boundaries and lower recombination rates in CZTS; studies of the optoelectronics of individual grain boundaries of known misorientation will be necessary to demonstrate that this is the case.

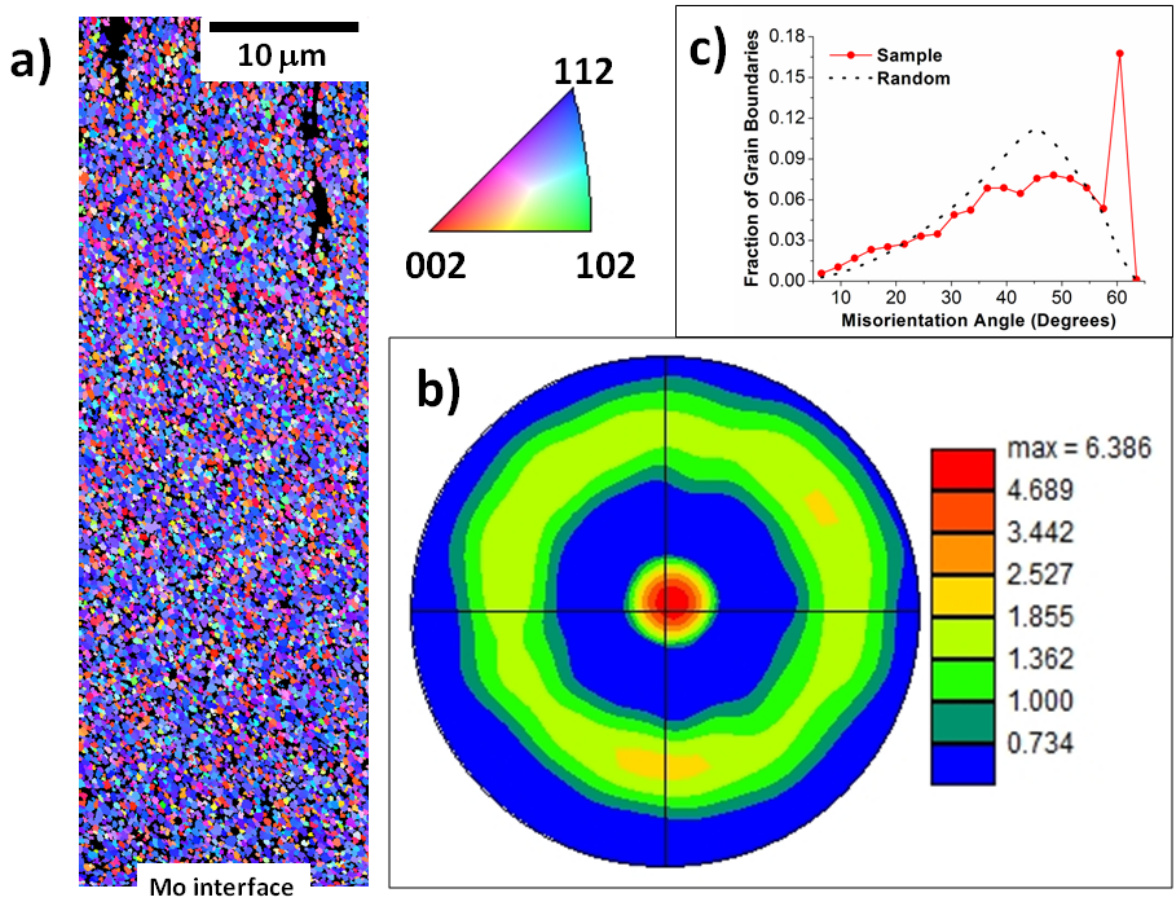

**Figure 6 -** EBSD analysis of an FIB-prepared inclined plane through a CZTS on Mo-coated SLG sample annealed at 600°C. (a) Orientation map along an inclined plane spanning from near the CZTS/Mo interface (bottom) to near the surface of the film (top). Because of the grazing angle of the FIB cut, the image shows the lateral dimensions of the grains as a function of depth in the film. The predominance of blue grains shows the dominant (112) texture as indicated by the grain orientation color key. (b) Pole figure for the ⟨112⟩ direction showing out-of-plane fiber texture six times random and some indications of weak in-plane texture. (c) Grain boundary misorientation angle distribution. Evidence of significant enrichment of Σ3 (60°) twin boundaries is apparent.

**CONCLUSIONS**

In this work we investigated the properties of $Cu_2ZnSnS_4$ (CZTS) photovoltaic absorber thin films synthesized using cosputtering from binary targets followed by annealing in sulfur vapor supplied in a two-zone furnace. XRD, Raman, and optical absorption indicate that the films are primarily kesterite, while Raman scattering indicates the presence of some ZnS. Lateral grain sizes in the 0.5 µm to 1 µm range are demonstrated using annealing temperatures of 550°C to 600°C. We explicitly investigated the effects of Na on grain growth in CZTS and demonstrated that Na can be added from sources other than SLG, which could enable processing on alternative substrates. Structural characterization using XRD and EBSD reveal (112) fiber texture in the films with significant enhancement of low-angle and $\Sigma 3$ grain boundaries, which may be beneficial in solar cells because of lower recombination rates.

**ACKNOWLEDGEMENTS**

This research was supported by the U.S. Department of Energy, Office of Basic Energy Sciences, Division of Materials Sciences and Engineering under Award DE-SC0001630.

**REFERENCES**

[1] T.M. Friedlmeier, H. Dittrich, and H.W. Schock, Ternary and Multinary Compounds, ed. by R.D. Tomlinson, A.E. Hill, and R.D. Pilkington (CRC Press, Boca Raton, 1998), p. 345.

[2] V. Fthenakis, Renew. Sust. Energ. Rev. 13, 2746 (2009).

[3] K. Zweibel, Science 328, 699 (2010).

[4] M.A. Green, Prog. Photovoltaics 14, 743 (2006).

[5] M.A. Green, Prog. Photovoltaics 17, 347 (2009).

[6] C. Wadia, A.P. Alivisatos, and D.M. Kammen, Environ. Sci. Technol. 43, 2072 (2009).

[7] H. Katagiri, K. Jimbo, W.S. Maw, K. Oishi, M. Yamazaki, H. Araki, and A. Takeuchi, Thin Solid Films 517, 2455 (2009).

[8] H. Katagiri, K. Jimbo, S. Yamada, T. Kamimura, W.S. Maw, T. Fukano, T. Ito, and T. Motohiro, Appl. Phys. Exp. 1, 041201 (2008).

[9] D.B. Mitzi, O. Gunawan, T.K. Todorov, K. Wang, and S. Guha, Sol. Energ. Mater. Sol. Cells 95, 1421 (2011).

[10] M.A. Green, K. Emery, Y. Hishikawa, and W. Warta, Prog. Photovoltaics 19, 84 (2011).

[11] K. Wang, O. Gunawan, T. Todorov, B. Shin, S.J. Chey, N.A. Bojarczuk, D. Mitzi, and S. Guha, Appl. Phys. Lett. 97, 143508 (2010).

[12] T.K. Todorov, K.B. Reuter, and D.B. Mitzi, Adv. Mater. 22, E156 (2010).

[13] J.J. Scragg, P.J. Dale, and L.M. Peter, Electrochem. Commun. 10, 639 (2008).

[14] H. Katagiri, Thin Solid Films 480, 426 (2005).

[15] P.A. Fernandes, P.M.P. Salome, and A.F. Da Cunha, Semicond. Sci. Technol. 24, 105013 (2009).

[16] H. Nukala, J.L. Johnson, A. Bhatia, E.A. Lund, W.M.H. Oo, M.M. Nowell, L.W. Rieth, and M.A. Scarpulla, Materials Research Society Symposium Proceedings, ed. D. Friedman, M. Stavola, W. Walukiewicz, and S. Zhang, Vol. 1268 (MRS, Warrendale, PA, 2010), EE3.4.

[17] J.L. Johnson, H. Nukala, E.A. Lund, W.M.H. Oo, A. Bhatia, L.W. Rieth, and M.A. Scarpulla, Materials Research Society Symposium Proceedings, ed. D. Friedman, M. Stavola, W. Walukiewicz, and S. Zhang, Vol. 1268 (MRS, Warrendale, PA, 2010), EE3.3.

[18] J.S. Seol, S.Y. Lee, J.C. Lee, H.D. Nam, and K.H. Kim, Sol. Energ. Mater. Sol. Cells 75, 155 (2003).

[19] H. Yoo and J. Kim, Sol. Energ. Mater. Sol. Cells 95, 239 (2011).

[20] S. Schorr, A. Weber, V. Honkimäki, and H.-W. Schock, Thin Solid Films 517, 2461 (2009).

[21] H. Katagiri, K. Saitoh, T. Washio, H. Shinohara, T. Kurumadani, and S. Miyajima, Sol. Energ. Mater. Sol. Cells 65, 141 (2001).

[22] T. Todorov and D.B. Mitzi, Direct liquid coating of chalcopyrite light-absorbing layers for photovoltaic devices. Eur. J. Inorg. Chem. 1, 17 (2010).

[23] P.A. Fernandes, P.M.P. Salome, and A.F. da Cunha, Thin Solid Films 517, 2519 (2009).

[24] P.K. Sarswat, M.L. Free, and A. Tiwari, Phys. Stat. Sol. (b), in press (2011).

[25] U. Rau, M. Schmitt, F. Engelhardt, O. Seifert, J. Parisi, W. Riedl, J. Rimmasch, and F. Karg, Solid State Commun. 107, 59 (1998).

[26] D. Braunger, D. Hariskos, G. Bilger, U. Rau, and H.W. Schock, Thin Solid Films 361, 161 (2000).

[27] T. Nakada, D. Iga, H. Ohbo, and A. Kunioka, Jpn. J. Appl. Phys. 36, 732 (1997).

[28] A. Rockett, Thin Solid Films 480–481, 2 (2005).

[29] D. Rudmann, G. Bilger, M. Kaelin, F.J. Haug, H. Zogg, and A.N. Tiwari, Thin Solid Films 431–432, 37 (2003).

[30] D. Abou-Ras, S. Schorr, and H.W. Schock, J. Appl. Crystallogr. 40, 841 (2007).

[31] D. Abou-Ras, U. Jahn, M. Nichterwitz, T. Unold, J. Klaer, and H.W. Schock, J. Appl. Phys. 107, 014311 (2010).

[32] D. Abou-Ras, C.T. Koch, V. Kastner, P.A. van Aken, U. Jahn, M.A. Contreras, R. Caballero, C.A. Kaufmann, R. Scheer, T. Unold, and H.W. Schock, Thin Solid Films 517, 2545 (2009).

[33] M.M. Nowell, M.A. Scarpulla, A.D. Compaan, X. Liu, D. Kwon, and K.A. Wieland, Conference Record of the 37th IEEE Photovoltaic Specialists Conference, Seattle, 2011.